\documentclass[oldversion,letterpaper,a4paper]{aa}
\usepackage{txfonts}
\usepackage{graphicx}
\ignorespaces

\begin{document}

\title{APECS - The Atacama Pathfinder Experiment Control System}

\author{D.\ Muders\inst{1} \and
        H.\ Hafok\inst{1} \and
        F.\ Wyrowski\inst{1} \and
        E.\ Polehampton\inst{1,2} \and
        A.\ Belloche\inst{1} \and \\
        C.\ K\"onig\inst{1} \and
        R.\ Schaaf\inst{1,3} \and
        F.\ Schuller\inst{1} \and
        J.\ Hatchell\inst{1,4} \and
        F.\ v.d.Tak\inst{1,5}}
\offprints{D.\ Muders, dmuders@mpifr-bonn.mpg.de}
\institute{Max-Planck-Institut f\"ur Radioastronomie, Auf dem H\"ugel 69, D-53121 Bonn, Germany \and
Space Science and Technology Department, Rutherford Appleton Laboratory, Chilton, Didcot, Oxfordshire, OX11 0QX, UK \and
Argelander-Institut f\"ur Astronomie, Universit\"at Bonn, Auf dem H\"ugel 71, D-53121 Bonn, Germany \and
School of Physics, University of Exeter, Stocker Road, Exeter EX4 4QL, UK\ \and
SRON National Institute for Space Research, Landleven 12, 9747 AD Groningen, The Netherlands}

\date{Received 4 April 2006 / Accepted 3 May 2006}

\abstract{
APECS is the distributed control system of the new Atacama Pathfinder
EXperiment (APEX) telescope located on the Llano de Chajnantor at an altitude
of 5107~m in the Atacama desert in northern Chile. APECS is based on Atacama
Large Millimeter Array (ALMA) software and employs a modern, object-oriented
design using the Common Object Request Broker Architecture (CORBA) as the
middleware. New generic device interfaces simplify adding instruments to the
control system. The Python based observer command scripting language allows
using many existing software libraries and facilitates creating more complex
observing modes. A new self-descriptive raw data format (Multi-Beam FITS or
MBFITS) has been defined to store the multi-beam, multi-frequency data.
APECS provides an online pipeline for initial calibration, observer feedback
and a quick-look display. APECS is being used for regular science observations
in local and remote mode since August 2005.}


\maketitle

\section{\label{introduction}Introduction}

Modern radio observatories such as the
new APEX\footnote{APEX is a collaboration between the Max-Planck-Institut
f\"ur Radioastronomie, the European Southern Observatory, and the Onsala
Space Observatory} submillimeter telescope (\cite{Guesten})
need complex control software to coordinate the
various hardware systems for the desired observations. The individual
instrument control computers and auxiliary devices like
synthesisers, etc.\ are typically distributed among different
locations throughout the observatory so that network communication
is essential. Real-time calculations are necessary to track the
target positions and handle observing patterns.
Monitoring hardware properties and environmental
conditions is important.

Since APEX is an experimental project, it will feature numerous bolometer
cameras and heterodyne array receivers operating in the atmospheric windows
between 150~GHz and 1.5~THz. These frontends are complemented by a set of
different continuum and spectral line backends to analyse the signals.
Frontends and backends can be connected to each other in many different ways for
observing. The APEX control system (APECS, \cite{Muders1})
therefore needs to be flexible to
handle the many different instruments and their combinations
and it must be easily extensible to include new devices. As a consequence of
the instrument complexity, the data formats for
the raw and calibrated data must be able to store all setup details.

APECS needs to provide
the standard radio observing modes like pointing, skydip, on-off
integrations or on-the-fly mapping.
Since observations at
submillimeter wavelengths are strongly affected by atmospheric absorption,
new calibration and observing modes need to be tried out and the control
software must support testing and implementing them in a simple fashion.
Radio astronomers also often wish to use a scripting language to create
observing macros. An online pipeline is needed
for feedback concerning the calibrations and for a quicklook
display of the scientific data. Finally, the location of APEX at
a very high site requires remote observing capabilities right from the
beginning.

In this letter we describe the design choices for APECS to fulfill the above
requirements and we highlight the most important parts of the software
developments that were made at the Max-Planck-Institut f\"ur Radioastronomie
over the course of several years until
now. The software has been iterated using user feedback that was collected
during the commissioning and early science observing phase of the APEX
telescope in 2004. Overall, APECS now provides a fully featured single-dish
telescope observing system that has been used by staff and visiting astronomers
for regular science observations since August 2005.

\section{APECS Design \& Implementation}

The APECS design phase began with evaluating existing radio telescope control
systems with the aim of re-using as much software as possible in order
to save manpower. The systems under consideration were, however, often
tightly connected to certain obsolete hardware
choices or difficult to maintain without the special knowledge of
the original developers.

Since APEX is a copy of one of the ALMA prototype antennas that were tested at
the VLA site in New Mexico, we closely followed the developments of the
initial ALMA control software. Although a prototype, it already showed
the modern development approach using a common network communication protocol
aimed at a flexible and maintainable software. The identical
telescope hardware interfaces of APEX and the ALMA antennas
allowed immediate application of the real-time
software which is usually very difficult to develop and test.
In addition, the participation of some of us in the then and current ALMA
software developments helped to master the initial learning curve quite
fast and allowed for some influence on the software design.

For the above reasons we decided to re-use the emerging ALMA software.
APECS is thus based on the framework of the so-called ALMA Common
Software (ACS, \cite{Raffi}) which uses the Common Object Request
Broker Architecture (CORBA, \cite{OMG}), an industry standard to provide a
multi-language, vendor-independent network communication layer. ACS delivers
the infrastructure for representing hardware devices in software via
distributed objects, i.e.\ software components that can be deployed
close to the hardware but accessed in the same manner from anywhere
in the control system without having to know the implementation details.
In addition, ACS provides means for automatic monitoring and logging.

APECS also re-uses parts of the ALMA Test Interferometer Control Software
(TICS, \cite{Glendenning}) which mainly provides the real-time system for basic
antenna control including astronomical coordinate system handling and observing
pattern elements.
TICS being a prototype software, we needed to improve its stability in some
areas for the normal science operations. This was done in collaboration with
some of the ALMA developers so that the effort was relatively small.

The ACS and TICS packages fulfill the requirements of common network
communication, automatic monitoring, real-time tracking and remote observing.
However, the overarching software to
use all hardware devices in a coordinated way necessary for astronomical
observations was only rudimentarily implemented in TICS because it aimed
only at testing the ALMA prototype antenna performance.
It was not suitable for operating a telescope\footnote{For ALMA this
functionality is currently under development but its time scale does not
match with the APEX schedule and it will not support multi-beam array
receivers.}. We therefore needed to develop that part of the software ourselves.

We began the development by defining the generic instrument and device
interfaces (cf.\ section \ref{generic_interfaces}) and the raw data format
interface (cf.\ section \ref{MBFITS}) since these areas needed to be stable
early on.
Subsequently, we developed the observer interface which provides a
scripting language for observing (cf.\ section \ref{obs_interface}),
the so-called Observing Engine to coordinate all hardware devices and
software tasks, the raw data writer (cf.\ section \ref{raw_data_writer})
and the online calibrator pipeline (cf.\ section \ref{data_calibrator}).

\begin{figure*}
\centering
{\rotatebox[origin=br]{0}{\includegraphics[width=17cm]{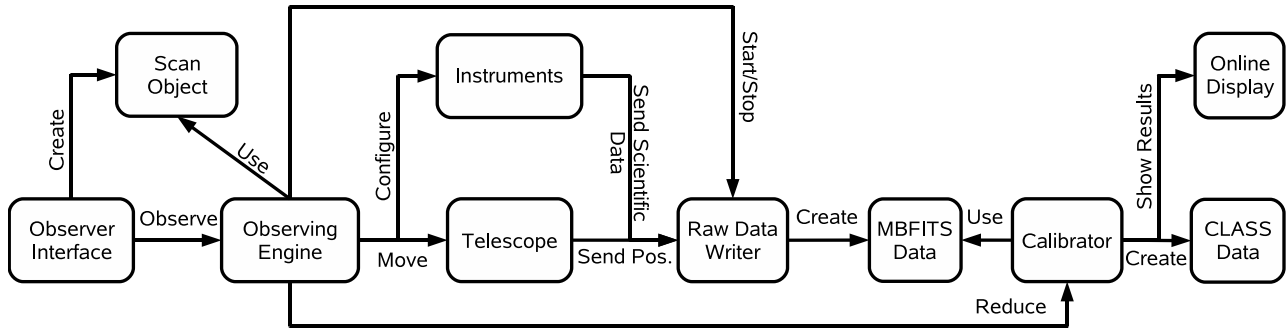}}}
\caption{This diagram illustrates the APECS observing pipeline structure.
The astronomer submits a request for a scan -- encoded as a so-called
Scan Object -- to the Observing Engine which then coordinates all hardware
and software tasks to perform the observation. It sets up the instruments,
moves the telescope to the desired position and starts the data recording.
The Raw Data Writer collects the data streams and creates an MBFITS file.
After each subscan the Calibrator provides calibrated data and shows results
on the online display for user feedback.
}
\label{apecs_pipeline}
\end{figure*}

These APECS core components are organised as a pipeline system (see
fig.\ \ref{apecs_pipeline}).
Observations are defined using so-called Scan Objects which contain the
full description of the next observation, i.e.\
the instrument setup details, target coordinate information and the
desired observing patterns.
The Scan Objects, that are created by the observer command line interface,
are sent to the Observing Engine which sets up all
necessary devices, controls the data acquisition and triggers
the online data calibration, reduction and display.

Aside from this main pipeline, we also developed a generic graphical
monitoring tool to view any system property and its alarm states and an
automatic observation logger that simplifies keeping detailed records
of the target, instrument setup and observing mode used for each scan.

Most of the APECS applications are written in
Python (\cite{Rossum}) which was chosen because a scripting language is
required for the user interfaces for observing and offline calibration, and
because ACS provides means to access the middleware from this language.
We also use Python for the non-interactive processes since
its high-level structures allow for a very efficient development of the
complex bookkeeping needed for instrument and pattern setups.
In addition,
there are Python wrappers to many existing compiled libraries to perform
heavy duty numerical calculations such as the atmospheric calibration or
processing the high raw data rates of up to several MB/s. Overall, we
believe that these advantages outweigh possible problems like the
dynamic typing.

In the following sections we highlight some details of the most important pieces
of the APECS developments.

\subsection{\label{generic_interfaces}Generic Instrument Interfaces}

One of the most important initial steps in a software development project is to
define the structure of packages and their interfaces. In a telescope control
system there is an additional need for interfaces to all the hardware devices
that are being used for the observations. We therefore began to collect
information about typical setups at other radio observatories to
eventually define a set of common instrument properties and methods
(\cite{Muders4}, \cite{Muders2}).

The important design decision was to require
that instruments of the same kind (e.g.\
heterodyne receivers, spectral backends, etc.) must 
{\em all} use the same high-level interface. This simplifies the setup for the
high-level observing software enormously because one merely adds a new
instrument name without having to worry about adding new features at
that level.

The implementation of these generic interfaces using the CORBA middleware
requires generic, though quite complex C++ code. We use a modified
version of a code generator originally developed at the U Bochum
(R.\ Lemke {\it priv. comm.}) to automatically create these program files.

The hardware side of these instrument interfaces is often served by
very simple computers such as micro-controllers which are not capable
of running the quite large middleware code directly. Instead, we employ
a simple text protocol following the SCPI \footnote{Standard Commands for
Programmable Instrumentation, SCPI Consortium, http://www.scpiconsortium.org}
standard (\cite{Hafok}).

\subsection{\label{MBFITS}MBFITS Raw Data Format}

In addition to the hardware interfaces, one also needs to determine
the data product interfaces early on. There was a lack of modern
single dish
raw data data format descriptions when the APECS developments began.
We therefore defined a new data format called Multi-Beam FITS (MBFITS,
\cite{Muders3}) to store the raw APEX data.

The MBFITS format was derived structurally from the ALMA
Test Interferometer FITS (ALMA-TI FITS, \cite{Lucas}) raw data
format, although a number of changes had to be made to accommodate the
special needs of the APEX and also the IRAM 30m and Effelsberg
100m telescopes where MBFITS is being used.

The MBFITS format uses the FITS standard (\cite{Wells}) and the
World Coordinate System (\cite{Greisen}) representation.
MBFITS is based on the scan-subscan-integration scheme
used by ALMA-TI fits and retains many of its keywords. However, due
to the changes in structure and additional keywords needed to
accommodate single-dish configurations, particularly multiple beam
observing and multiple frontend/backend combinations, the MBFITS
format can now be considered to be an independent format.

For each level of time granularity (scan, subscan, backend integration)
there are FITS binary tables to store the corresponding data. A special
monitoring table allows to record important instrument parameters
in parallel to the backend data stream for later analysis.

\subsection{\label{obs_interface}Observer Interface}

The main observer interface is implemented as a command line
interface (CLI) in a
Python interpreter thus fulfilling the user requirement to have
scripting with all options of a full programming
language. The observing commands have
been grouped according to functionality areas into catalog, target,
instrument, calibration, pattern and switch mode (e.g.\ wobbling
or frequency switching) setups.  We intentionally
implemented first a CLI to facilitate user scripting.
The future graphical user interface will use the existing commands.

In general, the commands are designed to be similar to those
found at other radio observatories. However, the typical setup of
frontend-backend chains is simplified due to the use of Python's
object-oriented features where we represent each instrument
by an object whose methods are used for further setup. This is
illustrated in the following example script to set up a 15 second
on-off observation of the CO~7--6 line in Orion-KL with a sky
reference 1,800$''$ to the east using the
FLASH 810~GHz receiver (\cite{Heyminck}) connected to one of the FFTS
spectrometers (\cite{Klein}) in a configuration with 8192 spectral
channels:

\vskip5pt
{
\small\tt
source 'orion-kl'

frontends 'flash810'

flash810.line 'CO(7-6)'

flash810.backends 'ffts1'

ffts1 numchan=8192

reference 1800,0

on 15
}

\subsection{\label{raw_data_writer}Raw Data Writer}

The Raw Data Writer must collect the telescope positions,
the backend data and the instrument configuration
and write them to an MBFITS file.
This is accomplished by a set of internal pipelines.
Each backend that is selected for a scan is associated with a Backend
Pipeline that receives the backend data, processes it,
and writes it to the corresponding MBFITS binary tables.

The so-called Monitor Pipeline receives telescope (and in the future
wobbler) position data, passes the data to the backend pipelines
for interpolation, and also writes it to the file. In addition,
it collects and writes user-defined monitor points from different devices.

\subsection{\label{data_calibrator}Data Calibrator}

The Data Calibrator (\cite{Polehampton}) provides calibration,
initial reduction, and display of data for both heterodyne and bolometer
receivers. This includes feedback to the
observing system for the basic pointing and focus observing modes.
The reduction proceeds on a subscan-by-subscan basis, retaining entities
that are required for further processing (e.g. references, calibrations).

Heterodyne calibration to the T$_{\rm A}^{\rm *}$ temperature scale is
carried out using an extended version of the
standard radio astronomy chopper wheel technique on sky, hot and cold loads.
The atmospheric calibration is calculated with the ATM model
(\cite{Pardo}) using the full Planck equation.
The final calibrated spectra are written to disk in
the CLASS\footnote{The Gildas software. http://www.iram.fr/IRAMFR/GILDAS}
format.
Bolometer data reduction is
carried out using libraries of The Bolometer Data Analysis Project (BoA,
\cite{Bertoldi}).

An offline command line interface based on the Python interpreter is
also provided for heterodyne data reduction. It uses exactly the same
methods as for the online system.

\section{Deployment}

The APECS software is deployed at three main locations: the telescope itself,
the control room at 5107~m altitude\footnote{Operating hard disks at such a
high altitude is technologically
challenging due to the low air pressure that can lead to head crashes.
APECS uses specially selected SCSI disks.}
on Chajnantor and the control room in the
APEX base camp in Sequitor near San Pedro de Atacama which is connected to
the mountain via a 32 Mbps microwave link.

Three main servers
provide the APECS pipeline system including the distributed objects
representing the hardware, the Observing Engine, the Raw Data Writer
and the Calibrator. A number of client stations are used
for local or remote observations from the high site or the base.
Remote observing from the partner institutes in Europe is possible
and has already been used.

One important aspect of the APECS software is the deployment in simulation
mode on a single computer without the need for any real instrument hardware.
This allows to test new developments in an end-to-end
fashion exactly as if performed at the telescope.

\section{Conclusion}

APECS is a modern, object-oriented telescope control system based on
the ALMA software framework using CORBA as the middleware. Its generic
interface approach
greatly simplifies adding new instruments. The automatic monitoring of
instrument properties facilitates debugging hardware problems. The
user-friendly, Python-based
scripting language that is employed for observations and data calibration
allows using many existing software libraries, thus
saving much development time. New observing modes can be easily added
at the scripting level. The new
MBFITS raw data format provides a self-descriptive, self-contained way
of storing all data that are necessary for further processing.
The online data processing pipeline provides calibrated spectra and
feedback for typical calibration scans. Overall, APECS is now a
mature telescope control system that can handle existing and
planned instruments and their data rates and has
the potential for future extensions.

\begin{acknowledgement}
We would like to thank the ALMA software developers who helped us
a lot to understand their software so that we could begin our own
developments within that framework.
\end{acknowledgement}

\end{document}